\documentclass{article}
\usepackage{spconf,amsmath,epsfig}

\usepackage{bm}
\usepackage{latexsym}
\usepackage{amsmath}
\usepackage{amssymb}

\def\bw{\mathbf{w}}

\def\bv{\mathbf{v}}

\def\bh{\mathbf{h}}

\title{Cooperative Jamming for Wireless Physical Layer Security}

\name{Lun Dong$^\dag$, Zhu Han$^\ddag$, Athina P. Petropulu$^\dag$
and H. Vincent Poor$^*$}
\address{$^\dag$Electrical \& Computer Engineering Department, Drexel University\\
$^\ddag$Electrical \& Computer Engineering Department, University of Houston \\
$^*$School of Engineering and Applied Science, Princeton University
\thanks{This research was supported by the
National Science Foundation under Grants CNS-06-25637, CCF-07-28208,
CNS-0831371 and CNS-0910461.}}

\begin{document}
%
\maketitle
\begin{abstract}
Cooperative jamming is an approach that has been recently proposed
for improving physical layer based security for wireless networks in
the presence of an eavesdropper. While the source transmits its
message to its destination, a relay node transmits a jamming signal
to create interference at the eavesdropper. In this paper, a
scenario in which the relay is equipped with \emph{multiple}
antennas is considered. A novel system design is proposed for
determining the antenna weights and transmit power of source and
relay, so that the system secrecy rate is maximized subject to a
total transmit power constraint, or, the transmit power is minimized
subject to a secrecy rate constraint. Since the optimal solutions to
these problems are difficult to obtain, suboptimal closed-form
solutions are proposed that introduce an additional constraint,
i.e., the complete nulling of jamming signal at the destination.
\end{abstract}
\section{Introduction}
\label{sec:intro}

Security is an important issue in wireless networks due to the open
wireless medium. Security against an eavesdropper is typically
achieved via cryptographic algorithms that are implemented at higher
network layers \cite{Sklavos}. Physical (PHY) layer based security,
a line of work that has attracted considerable recent attention in
this context, exploits the physical characteristics of the wireless
channel to transmit messages securely (see \cite{LiangPoorSham} for
a review of recent developments in this area). The idea was
pioneered by Wyner, who introduced the wiretap channel and
established the possibility of creating perfectly secure
communication links without relying on private (secret) keys
\cite{Wyner}. Wyner showed that when an eavesdropper's channel is a
degraded version of the main channel, the source and destination can
exchange perfectly secure messages at a non-zero rate, while the
eavesdropper can learn almost nothing about the messages from its
observations. The maximal rate of perfectly secret transmission from
the source to its intended destination is named the \emph{secrecy
rate}. However, the feasibility of traditional PHY layer security
approaches based on single antenna systems is hampered by channel
conditions: absent feedback, if the channel between source and
destination is worse than the channel between source and
eavesdropper, the secrecy rate is typically zero.

Some recent work has been proposed to overcome this limitation by
taking advantage of user cooperation
\cite{Dong:Allerton}-\cite{Lai}. For conventional wireless networks
without secrecy constraints (i.e., without eavesdropper), the most
common strategy for user cooperation is \emph{cooperative relaying},
e.g., the well-known decode-and-forward and amplify-and-forward
schemes. Cooperative relaying with secrecy constraints was discussed
in \cite{Dong:Allerton}-\cite{Aggarwal}. Cooperative jamming is
another approach to implement user cooperation for wireless networks
with secrecy constraints \cite{Negi}-\cite{Lai}. In cooperative
jamming, a relay transmits a jamming signal at the same time when
the source transmits the message signal, with the purpose of jamming
the eavesdropper. Existing works on cooperative jamming have focused
primarily on the case of one single-antenna relay, and on the
analysis of secrecy rate and the capacity-achieving strategy.

In this paper, we consider a different system model and different
design objectives for implementing cooperative jamming. We consider
a scenario in which a source communicates with a destination in the
presence of one eavesdropper. The communication is aided by a relay
that is equipped with \emph{multiple} antennas that provide more
degrees of freedom for the relay channel. Our goal is to assign
weights to the antenna elements in an optimum fashion, and also
allocate the power of source and relay in an optimum fashion. The
weight and power design problem is formulated as the following
optimization problem: (1) maximize achievable secrecy rate subject
to a total transmit power constraint; or (2) minimize the total
transmit power subject to a secrecy rate constraint. We assume that
global channel state information (CSI) is available for system
design. As the optimum design is in general difficult, we consider a
suboptimal weight design, i.e., to completely null out the jamming
signal at the destination. From the simulation results, cooperative
jamming could significantly improve the system performance
especially when the eavesdropper is close to the relay.

\section{System model and problem formulation}
\label{sec:model}

We consider a wireless network model consisting of one source, one
trusted relay\footnote{We still adopt the name ``relay'', though it
is not used for relaying here.}, one destination, and one
eavesdropper. The source, destination and eavesdropper are each
equipped with a single omni-directional antenna each, while the
relay has $N>1$ omni-directional antennas\footnote{This can also be
understood as multiple relays with one antenna each.}. All channels
are assumed to undergo flat fading. We denote by $h_{SD}$ the
source-destination channel, by $h_{SE}$ the source-eavesdropper
channel, by $\bh_{SR}$ the source-relay channel ($N \times 1$ column
vector), by $\bh_{RD}$ the relay-destination channel ($N \times 1$
column vector), and by $\bh_{RE}$ the relay-eavesdropper channel ($N
\times 1$ column vector). A narrowband message signal $s$ is to be
transmitted from the source to the destination. The power of the
message signal $s$ is normalized to unity, i.e, $\mathbb{E}\{|s|^2\}
=1$. The total power budget for transmitting the message signal to
its destination is $P$. Thermal noise at any node is assumed to be
zero-mean white complex Gaussian with variance $\sigma^2$.

\subsection{Cooperative jamming}

In cooperative jamming, while the source transmits, the relay
transmit a jamming signal that is independent of the source message.
The goal is to interfere with the eavesdropper's received signal.
More specifically, the source transmits the message signal
$\sqrt{P_s} s$, where $P_s$ is the transmit power of the source; at
the same time, the relay transmits a weighted version of a common
jamming signal $z$, i.e., $\bw z$, where $\bw$ is the weight vector
applied on the $N$ antennas. The transmit power budget for
transmitting the jamming signal  is thus $P_j = P - P_s$.

The received signal at the destination equals
\begin{eqnarray} \label{SigDesCJ}
y_d &=& \sqrt{P_s} h_{SD} s + \bw^\dag \bh_{RD} z + n_d \ ,
\end{eqnarray}
where $n_d$ represents white complex Gaussian noise at the
destination and $(\cdot)^\dag$ represents the Hermitian transpose.
The received signal at the eavesdropper equals
\begin{eqnarray} \label{SigEavCJ}
y_e &=& \sqrt{P_s} h_{SE} s + \bw^\dag \bh_{RE} z + n_e \ ,
\end{eqnarray}
where $n_e$ represents white complex Gaussian noise at the
eavesdropper.

\subsection{Problem formulation} \label{probForm}
In the presence of an eavesdropper, \emph{secrecy rate} is the
figure of merit to represent the maximal \emph{secrecy information}
rate one can transmit from a source to its destination. Recall that
the secrecy rate is $R_s = \max\left\{0,R_d- R_e \right\}$ where
$R_d$ is the rate from the source to the destination and $R_e$ is
the rate from the source to the eavesdropper \cite{Wyner}. We
consider the practical case in which the system can be designed so
that the secrecy rate is positive. In that case, the secrecy rate
can be simplified to $R_s = R_d-R_e$. From (\ref{SigDesCJ}) and
(\ref{SigEavCJ}), the secrecy rate is given by
\begin{eqnarray} \label{SecCapCJ}
R_s &=& \log_2 \left(1 + \frac{P_s |h_{SD}|^2}{|\bw^\dag \bh_{RD}|^2
+ \sigma^2}\right) \nonumber \\
&& - \log_2 \left(1+ \frac{P_s |h_{SE}|^2}{|\bw^\dag \bh_{RE}|^2 +
\sigma^2}\right) \ .
\end{eqnarray}

In the subsequent section we determine the relay weights ${\bf w}$
and the power $P_s$ based on the objectives: (i) maximize secrecy
rate subject to a transmit power constraint $P_0$, or,  (ii)
minimize the transmit power subject to a secrecy rate constraint
$R_s^0$. We assume that global CSI is available for system design (a
common assumption in the PHY security literature), even the
eavesdropper's channel are known. Information on the eavesdropper's
channel can be obtained in cases where the eavesdropper is active in
the network and their transmissions can be monitored. This is
applicable particularly in networks combining multicast and unicast
transmissions, in which terminals play dual roles as legitimate
receivers for some signals and eavesdroppers for others.

\section{System Design}
\label{systemdesign}

In this section we address system design that maximizes the secrecy
rate subject to a total transmit power constraint $P_0$ or minimize
the total transmit power subject to a secrecy rate constraint
$R_s^0$. There are two aspects in system design: one is to determine
the optimal transmit power allocated to the source and to relay;
another is to design the optimal relay weights. In the following, we
first fix $P_s$ to obtain the weights for secrecy rate maximization
or power minimization, and then find the optimal value of $P_s$.

\subsection{Secrecy rate maximization} \label{secCapMax}
From (\ref{SecCapCJ}), it can be seen that the secrecy rate is a
product of two correlated generalized eigenvector problems and is in
general difficult to handle. To simplify the analysis, in the
following we consider a suboptimal design. We add one more
constraint to completely null out the jamming signal at the
destination, i.e., $\bw^\dag \bh_{RD} = 0$. Then, the problem of
secrecy rate maximization can be formulated as
\begin{eqnarray} \label{opt_CJ_single}
& \arg \max\limits_{\bw} \ |\bw^\dag \bh_{RE}|^2 &
\\ \nonumber & \mathrm{s.t.} \left\{\begin{array}{ll} \nonumber \bw^\dag \bh_{RD} = 0 \\
       \|\bw\|^2 \leq P_j \end{array} \right.
\end{eqnarray}
where $\|\cdot\|$ denotes the vector 2-norm. It is easy to show that
the inequality constraint (i.e., $\|\bw\|^2 \leq P_j$) in
(\ref{opt_CJ_single}) is equivalent to the equality constraint
$\|\bw\|^2 = P_j$ by contradiction.

The Lagrangian of (\ref{opt_CJ_single}) is
\begin{eqnarray}
L(\bw^\dag,\lambda,\eta) = (2 \bh_{RE}^\dag \bw) \bh_{RE} + \lambda
\bh_{RD} + 2 \eta \bw
\end{eqnarray}
where $\lambda$ and $\eta$ are Lagrange multipliers. As
$\bh_{RE}^\dag \bw$ is a scalar, by setting the Lagrangian to zero,
we can see that $\bw$ is a linear combination of $\bh_{RD}$ and
$\bh_{RE}$, represented as
\begin{eqnarray}
\bw = a \bh_{RD} + b \bh_{RE} \ .
\end{eqnarray}
Substituting $\bw$ into the two equality constraints in
(\ref{opt_CJ_single}), we can solve $a$ and $b$. Notice that the
values of $a$ and $b$ are not unique: it is straightforward to see
that, if $\bw^*$ is the solution of (\ref{opt_CJ_single}), after
rotating an arbitrary phase, $e^{j \theta} \bw^*$ is still the
solution of (\ref{opt_CJ_single}). A possible selection is $a = -
\sqrt{P_j} \mu \bh_{RD}^\dag\bh_{RE}$ and $b = \sqrt{P_j} \mu
\|\bh_{RD}\|^2$ where
\begin{eqnarray}
\mu = \left[\|\bh_{RD}\|^4 \|\bh_{RE}\|^2 - \|\bh_{RD}\|^2
|\bh_{RD}^\dag\bh_{RE}|^2\right]^{-1/2} \ .
\end{eqnarray}

Next, we find the optimal value of $P_s$. Noticing that the above
designed weight vector $\bw$ is proportional to $\sqrt{P_s}$, let us
denote $\bw$ by $\bw = \sqrt{P_s} \bv$ where
\begin{eqnarray}
\bv = (- \mu \bh_{RD}^\dag\bh_{RE}) \bh_{RD} + (\mu \|\bh_{RD}\|^2)
\bh_{RE} \ .
\end{eqnarray}
Substituting it into the expression of $R_s$, the secrecy rate can
be expressed as a function of $P_s$:
\begin{eqnarray} \label{Cap_Ps}
R_s(P_s) = \log_2\left(\frac{e_0 + e_1 P_s + e_2 P_s^2}{f_0 + f_1
P_s} \right)
\end{eqnarray}
where $e_i$ and $f_i$ are coefficients independent of $P_s$ and
given by
\begin{eqnarray}
e_0 & = & \sigma^2(\sigma^2 + P_0 |\bv^\dag \bh_{RE}|^2) \ , \\
e_1 & = & (|h_{SD}|^2 P_0 - \sigma^2) |\bv^\dag \bh_{RE}|^2 + |h_{SD}|^2 \sigma^2  \ , \\
e_2 & = & - |h_{SD}|^2 |\bv^\dag \bh_{RE}|^2 \ , \\
f_0 & = & \sigma^2(\sigma^2 + P_0 |\bv^\dag \bh_{RE}|^2) \ , \\
f_1 & = & \sigma^2(|h_{SE}|^2- |\bv^\dag \bh_{RE}|^2) \ .
\end{eqnarray}

Taking the derivative of $2^{R_s(P_s)}$ and setting to zero, the
optimal value of $P_s$ is the solution of the quadratic equation
\begin{eqnarray} \label{quadEq_CJ}
e_2 f_1 (P_s)^2 + 2 e_2 f_0 P_s + (e_1 f_0 - e_0 f_1) = 0 \ .
\end{eqnarray}
In case no solution exists within $(0,P_0]$, it holds that $P_s =
P_0$ (i.e., the case of direct transmission without jamming).

\subsection{Transmit power minimization}
We again consider to completely null out the jamming signal at the
destination. For a fixed $P_s$, the problem of secrecy rate
maximization can be formulated as
\begin{eqnarray}
& \arg \min\limits_{\bw} \ \|\bw\|^2 &
\\\ \nonumber & \mathrm{s.t.} \left\{\begin{array}{ll} \nonumber \bw^\dag \bh_{RD} = 0 \\
       |\bw^\dag \bh_{RE}|^2 \geq \rho \end{array}
       \right.
\end{eqnarray}
where $\rho = \frac{P_s |h_{RE}|^2}{2^{-R_s^0} (1+ P_s
|h_{RD}|^2/\sigma^2) -1} - \sigma^2$. By using the Lagrange
multiplier method (similarly as in Section \ref{secCapMax}), it can
be shown that $\bw$ is a linear combination of $\bh_{RD}$ and
$\bh_{RE}$, represented as $\bw = a \bh_{RD} + b \bh_{RE}$. A
possible selection is $a = - \mu \bh_{RD}^\dag\bh_{RE}$ and $b = \mu
||\bh_{RD}||^2$ where
\begin{eqnarray}
\mu = \sqrt{\frac{\rho}{\|\bh_{RD}\|^2 \|\bh_{RE}\|^2 -
|\bh_{RD}^\dag\bh_{RE}|^2}} \ .
\end{eqnarray}

Let us denote $\bw = \sqrt{\rho} \bv$ where
\begin{eqnarray}
\bv = \frac{- \bh_{RD}^\dag\bh_{RE} \bh_{RD} + \|\bh_{RD}\|^2
\bh_{RE}}{\|\bh_{RD}\|^2 \|\bh_{RE}\|^2 - |\bh_{RD}^\dag\bh_{RE}|^2}
\ .
\end{eqnarray}
Then, the total transmit power can be represented as
\begin{eqnarray} \label{totalpower}
P_s + \|\bw\|^2 = \frac{e_0 + e_1 P_s + e_2 P_s^2}{f_0 + f_1 P_s}
\end{eqnarray}
where
\begin{eqnarray}
e_0 & = & -(2^{-R_s^0} - 1)\sigma^2 \|\bv\|^2 \ , \\
e_1 & = & 2^{-R_s^0} - 1 + (|h_{SE}|^2 - 2^{-R_s^0} |h_{SD}|^2) \| \bv\|^2  \ , \\
e_2 & = & 2^{-R_s^0}|h_{SD}|^2/\sigma^2 \ , \\
f_0 & = & 2^{-R_s^0} - 1 \ , \\
f_1 & = & 2^{-R_s^0}|h_{SD}|^2/\sigma^2 \ .
\end{eqnarray}
Then, the optimal value of $P_s$ is obtained by solving the
quadratic equation which is of the same form as (\ref{quadEq_CJ}).

\section{Simulations}
\begin{figure}[thb]
 \centerline{\epsfig{figure=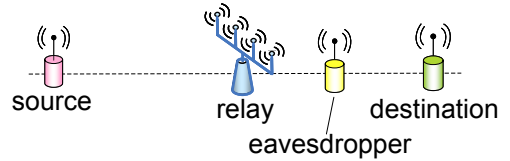,width=6.5 cm}}
\caption{Simulation model.}
 \label{simulationmodel}
\end{figure}

For convenience, we consider a simple one-dimension system model, as
illustrated in Fig. \ref{simulationmodel}, in which the source,
relay, destination and eavesdropper are placed along a line. To
highlight the effects of distances, channels are modeled by a simple
line-of-sight channel model including the path loss (path loss
exponent is $3.5$) and a random phase (uniformly distributed). The
source-destination distance is fixed at $50$ m, and the source-relay
distance is fixed at $25$ m (i.e., the relay is located at the
middle point of source and destination). The noise power $\sigma^2$
is $-100$ dBm.

 \vspace*{-0.1in}

\begin{figure}[t]
 \centerline{\epsfig{figure=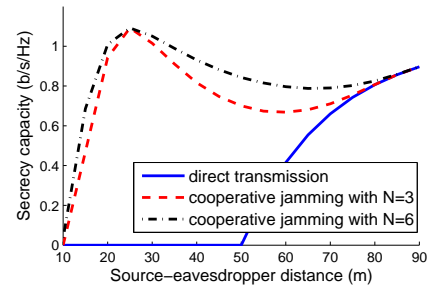,width=6.1cm}}
\caption{Secrecy rate versus source-eavesdropper distance.}
 \label{secrecyrate}
\end{figure}

The position of the eavesdropper is varied so that the
source-eavesdropper distance varies  from $10$ m to $90$ m. Fig.
\ref{secrecyrate} shows the secrecy rate versus the
source-eavesdropper distance for a transmit power constraint $P_0
\leq -40$ dBm. For direct transmission without the relay's help, the
secrecy rate is positive only if the source-eavesdropper distance is
larger than the source-destination distance. Also, the secrecy rate
of direct transmission increases with the increase of
source-eavesdropper distance, as the rate at the eavesdropper $R_j$
decreases. For cooperative jamming, the closer the eavesdropper is
located to the relay, the higher the secrecy rate is. This is
because when  relay and eavesdropper are close, even a small amount
of power allocated to the relay can create enough interference at
the eavesdropper, and a large amount of power can be used for the
source to transmit the message signal. When the eavesdropper moves
away from the relay and closer to the source, the secrecy rate
decreases, since more jamming power is needed for creating larger
interference and less power is available for the source to transmit
message signal. When the eavesdropper moves away from both the relay
and source, it is interesting to see that the secrecy rate at first
decreases, then increases, and eventually becomes equal to the
secrecy rate of direct transmission. This is because when the
eavesdropper is very far from the relay and the source, we should
spend most of the power on transmitting the message signal. In that
situation it is not worth spending a large amount of power on the
jamming signal, since the received power of the message signal at
the eavesdropper is always small (regardless of jamming) due to the
large path loss. Also, as expected, increasing the number of
antennas of the relay $N$ can always improve the secrecy rate.

\begin{figure}[t]
 \centerline{\epsfig{figure=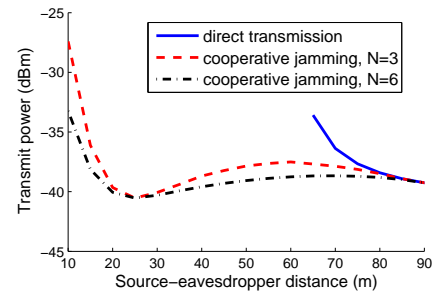,width=6.1cm}}
\caption{Transmit power versus source-eavesdropper distance.}
 \label{transmitpower} \vspace*{-0.1in}
\end{figure}

Fig. \ref{transmitpower} shows the total transmit power versus the
source-eavesdropper distance for a secrecy rate constraint $R_s^0
\geq 1$ b/s/Hz. For the curve corresponding to direct transmission,
we only show the feasible region in which the required secrecy rate
can be satisfied. The curves for cooperative jamming exhibits
similar characteristics to Fig. \ref{secrecyrate}, therefore a
detailed discussion is omitted.

\section{Conclusions}
In this paper, we have addressed practical system design problems
for the cooperative jamming protocol for secure wireless
communications in the presence of a relay with multiple antennas.
For cooperative jamming, while the the source transmits its message
signal, the relay transmits a jamming signal to create interference
at the eavesdropper. The multiple antennas at the relay can provide
degrees of freedom for the relay channel and thus eliminate the
effects of jamming signals at the destination. The objectives of our
system design are the secrecy rate maximization subject to a total
power constraint and the transmit power minimization subject to a
secrecy rate constraint. Simulation results show that cooperative
jamming could significantly improve the system performance
especially when the eavesdropper is close to the relay. Future work
includes the investigations of performance degradation and system
design in the presence of imperfect channel estimates.

\vspace*{-0.1in}

\end{document}